\documentclass[twocolumn,prl,letter]{revtex4}
\usepackage{amssymb}
\usepackage{graphicx}
\usepackage{natbib}

\begin{document}

\title{Conduction electron spin-lattice relaxation time in the MgB$_2$
superconductor}

\author{F. Simon,$^{\ast}$ F. Mur\'{a}nyi,$^{\dag}$ T. Feh\'er, A. J\'{a}nossy}
\affiliation{Budapest University of Technology and Economics, Institute of Physics\\
and Condensed Matter Research Group of the Hungarian Academy of\\
Sciences, H-1521 Budapest, PO BOX 91, Hungary}

\author{L. Forr\'o}
\affiliation{IPMC/SB Swiss Federal Institute of Technology (EPFL),
CH-1015 Lausanne-EPFL, Switzerland}
\author{C. Petrovic,$^{\ddag}$ S.L. Bud'ko, P.C. Canfield}

\affiliation{Ames Laboratory, U.S. Department of Energy and Department of Physics\\
and Astronomy, Iowa State University, Ames, Iowa 50011, USA}

\date{\today}

\begin{abstract}
The spin-lattice relaxation time, $T_{1}$, of conduction electrons
is measured as a function of temperature and magnetic field in
MgB$_2$. The method is based on the detection of the $z$ component
of the conduction electron magnetization under electron spin
resonance conditions with amplitude modulated microwave excitation.
Measurement of $T_{1}$ below $T_c$ at 0.32 T allows to disentangle
contributions from the two Fermi surfaces of MgB$_{2}$ as this field
restores normal state on the Fermi surface part with $\pi$ symmetry
only.

\end{abstract}

\maketitle




\section{Introduction}

The conduction electron spin-lattice relaxation time in metals,
$T_{1}$, is the characteristic time for the return to thermal
equilibrium of a spin system driven out of equilibrium by e.g. a
microwave field at electron-spin resonance (ESR)\ or a
spin-polarized current. The applicability of metals in
``spintronics" devices in which information is processed using
electron spins \cite{FabianRMP} depends on a sufficiently long spin
life-time. In
pure metals $T_{1}$ is limited by the Elliott mechanism \cite{Elliott,Yafet}%
, i.e. the scattering of conduction electrons by the random spin-orbit
potential of non-magnetic impurities or phonons. In superconductors, the
Elliott mechanism becomes ineffective and a long $T_{1}$ is predicted well below $%
T_{c}$ \cite{Yafet}. Here we report the direct measurement of the
spin-lattice relaxation time of conduction electrons\ in MgB$_{2}$
in the superconducting state. The motivation to study the magnetic
field and temperature dependence of $T_{1}$ is two-fold: i) to test
the predicted lengthening of $T_{1}$ to temperatures well below
$T_{c}$, ii) to measure the contributions to $T_{1}$ from different
Fermi surface sheets and to compare with the corresponding momentum
life-times, $\tau $.

The lengthening of $T_{1}$ has been observed in a restricted
temperature range below $T_{c}$ in the fulleride superconductor,
K$_{3}$C$_{60}$ by measuring the conduction electron-spin resonance
(CESR) line-width, $\Delta H$ \cite{NemesPRB}. This method
assumes $1/T_{1}=1/T_{2}=\gamma _{\text{e}}\Delta H$, where $\gamma _{\text{e%
}}/2\pi =28.0\text{ GHz/T}$ is the electron gyromagnetic ratio, and
$1/T_{2}$ is the spin-spin or transversal relaxation rate. It is
limited to cases where the homogeneous broadening of the CESR line
due to a finite spin lifetime outweighs $\Delta H_{\text{inhom}}$,
the line broadening from inhomogeneities of the magnetic field. In a
superconducting powder sample, the CESR line is inhomogeneously
broadened below the irreversibility line due to the distribution of
vorteces, which is temperature and magnetic field dependent. This
prevents the measurement of $T_{1}$ from the line-width and calls
for a method to directly measure $T_{1}$. Electron spin echo
techniques, which usually enable the measurement of $T_{1}$, are not
available for the required nanosecond time resolution range. The
magnetic resonance technique, termed longitudinally detected (LOD) ESR \cite%
{MuranyiJMR,SimonJMR} used in this work allows to measure $T_{1}$'s
as short as a few ns. The method is based on the measurement of the
electron spin magnetization along the magnetic field, $M_z$, using
modulated microwave excitation. $M_z$ recovers to its equilibrium
value with the $T_1$ time-constant, thus the method allows the
direct measurement of $T_{1}$ independent of magnetic field
inhomogeneities.

MgB$_{2}$ has a high superconducting transition temperature of
$T_{c}=39$ K
\cite{AkimitsuNAT} and its unusual physical properties \cite%
{BouquetPRL2001,SzaboPRL2001,BouquetPRL2002,ARPESPRL2003} are
attributed \cite{MazinPRL2001,LouieNAT2002} to its disconnected,
weakly interacting Fermi surface (FS) parts. The Fermi surface
sheets derived from B-B bonds with $\pi $ and $\sigma $ characters
($\pi $ and $\sigma $ FS) have smaller and higher electron-phonon
couplings and superconductor gaps, respectively, and contribute
roughly equally to the density of states (DOS). The strange band
structure leads to unique thermodynamic properties: magnetic fields
of about 0.3-0.4 T restore the $\pi $ FS\ well below $T_{c}$ for all
field orientations in polycrystalline samples but the material
remains superconducting to much higher fields. This is characterized
by a small and
nearly isotropic upper critical field, $H_{c2}^{\pi }\sim 0.3-0.4$ T \cite%
{BouquetPRL2002,EskildsenPRL2002} and a strongly anisotropic one, $%
H_{c2}^{\sigma }=2-16$ T,
\cite{SimonPRL2001,BouquetPRL2002,AngstPRL2002} related to the $\pi
$ and $\sigma $ Fermi surface sheets, respectively. Our
measurements at low fields and low temperatures determine $T_{1}$ from the $%
\pi $ FS alone, while high field and high temperature experiments measure $%
T_{1}$ averaged over the whole FS. We find that spin relaxation in high
purity MgB$_{2}$ is temperature independent in the high field normal state
between 3 K and 50 K, indicating that it arises from non magnetic
impurities. Spin relaxation times for electrons on the $\pi $ and $\sigma $
Fermi surface sheets are widely different but are not proportional to the
corresponding momentum relaxation times.

\section{Experimental}

The same MgB$_{2}$ samples were used as in a previous study \cite%
{SimonMgB2PRB2005}. Thorough grinding, particle size selection and
mixing with SnO$_{2}$, an ESR silent oxide, produced a fine powder
with well separated small metallic particles. The nearly symmetric
appearance of the CESR signal \cite{Dyson} proves that penetration
of microwaves is homogeneous and that the particles are smaller than
the microwave penetration depth of $\sim 1 \mu\text{m}$. SQUID
magnetometry showed that grinding and particle selection do not
affect the superconducting properties. The particles are not single
crystals but rather aggregates of small sized single crystals.
Continuous wave (cw) and longitudinally detected ESR experiments
were performed in a home-built apparatus \cite{SimonJMR} at 9.1 and
35.4 GHz microwave frequencies, corresponding to 0.32 and 1.27 T
resonance magnetic fields. The 9.1 GHz apparatus is based on a
loop-gap resonator with a low quality factor ($Q\sim 200$) and the
35.4 GHz instrument does not employ a microwave cavity at all. The
cw-ESR was detected using an audio frequency magnetic field
modulation. Line-widths are determined by Lorentzian fits to the
cw-ESR data. For the LOD-ESR, the microwaves are amplitude modulated
with $f=\Omega /2\pi $ of typically 10 MHz and the resulting varying
$M_{z}$ component of the sample magnetization is detected with a
coil which is parallel to the external magnetic field and is part of
a resonant circuit that is tuned to $f $ and is matched to 50 Ohms.
cw-ESR at 420 GHz (centered at 15.0 T) was performed at EPFL using a
quasi-optical microwave bridge with no resonant cavities.


\section{Results}

\subsection{Relaxation in the normal state}

\begin{figure}[tbp]
\includegraphics[width=0.8\hsize]{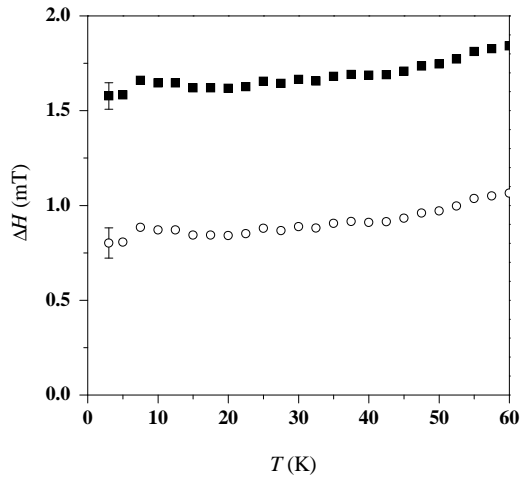}
\caption{CESR line-width of MgB$_2$ as a function of temperature for
the 15 T CESR measurement ($\blacksquare$). Open circles
($\bigcirc$) show the homogeneous line-width ($\Delta
H_{\text{hom}}$) after correcting for the field dependent broadening
as explained in the text. Representative error bars are shown at the
lowest temperature.} \label{15T_Tdep}
\end{figure}

The low temperature behavior of the spin-lattice relaxation time in
MgB$_{2}$ in the normal state can be measured using cw-ESR from the
homogeneous line-width, $\Delta H_{\text{hom}}$, using
$1/T_{1}=\gamma _{\text{e}}\Delta H_{\text{hom}}$ at high fields,
$H>H_{\text{c2}}$ that suppresses
superconductivity. The maximum upper critical field is $H_{\text{c2,max}%
}\sim 16$ T for particles with field in the $(a,b)$ crystallographic
plane in the polycrystalline sample at zero temperature
\cite{FinnemorePRL2001}. We did not observe any effects of
superconductivity on the CESR, at 15 T it is
suppressed in the full sample above a temperature of a few K. Fig. \ref%
{15T_Tdep} shows that the temperature dependence of the CESR
line-width at 15 T is small below 40 K.

\begin{figure}[tbp]
\includegraphics[width=0.8\hsize]{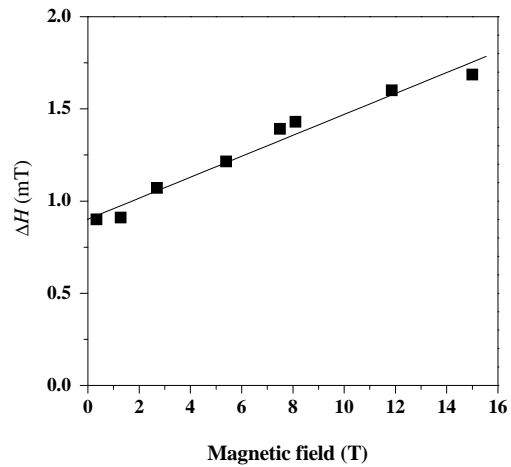}
\caption{ESR line-width of MgB$_2$ as a function of magnetic field
measured at 40 K ($\blacksquare$). Solid curve is a linear fit to
the data with parameters given in the text.} \label{Hdep}
\end{figure}

The CESR line-width is magnetic field dependent as shown in Fig.
\ref{Hdep} at 40 K: it is linear as function of magnetic field with
$\Delta H=\Delta H_{0}+b\ast
H$, where $\Delta H_{0}=0.90(1)$ mT is the residual line-width and $%
b=0.057(1)$ mT/T. The residual homogeneous line-width corresponds to
$T_{1}=6.3$ ns at 40 K. The linear relation can be used to correct
the 15 T CESR line-width data
to obtain the homogeneous contribution, $\Delta H_{\text{Hom}}(T)=\Delta H(%
\text{15 T},T)-15 \text{T}\cdot b$ as the magnetic field dependence
is expected to be
temperature independent. We show the homogeneous line-width in Fig. \ref%
{15T_Tdep}. We find that it is temperature independent within experimental
precision between 3 and 50 K. This means that the spin-lattice relaxation
time flattens to a residual value that is given by non-magnetic impurities.

\subsection{Relaxation in the superconducting state}

\begin{figure}[tbp]
\includegraphics[width=0.8\hsize]{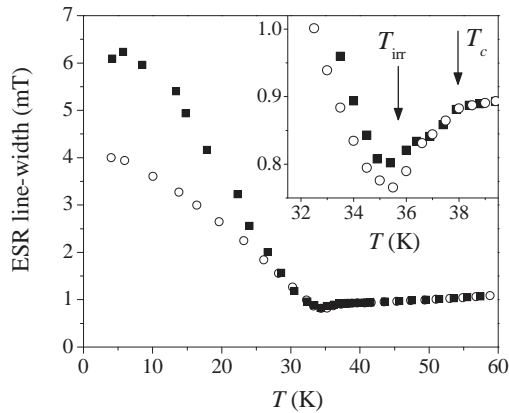}
\caption{Inhomogeneous CESR line broadening in MgB$_2$ below $T_c$
at 0.32 T. Full and open symbols show the CESR line-width for up and
down magnetic field sweeps, respectively. Inset shows the data near
$T_c$. Note the line narrowing between $T_c$ and $T_{\text{irr}}$
and the field sweep direction dependent line-widths below the
irreversibility temperature.} \label{ESR_linewidths}
\end{figure}



In type II superconductors, CESR arises from thermal excitations and
from normal state vortex cores. The inhomogeneity of the magnetic
field in the vortex lattice or glass states does not broaden the
CESR line. The local magnetic field inhomogeneity is averaged since\
within the spin life-time itinerant electrons travel long distances
compared to the inter-vortex distance \cite{NemesPRB}. This is in
contrast to the NMR case where the line-shape is affected: the
nuclei are fixed to the crystal and nuclei
inside and outside the vortex cores experience different local fields \cite%
{Pincus1964}. In other words, a superconducting a single crystal
sample would display a narrow conduction electron ESR line if there
were no irreversible effects. However, the CESR line is
inhomogeneously broadened below the irreversibility line for a
superconducting powder sample: the vortex distribution depends on a
number of factors such as the thermal and magnetic field history,
grain size and, for an anisotropic superconductor such as MgB$_{2}$,
on the crystal orientation with respect to the magnetic field also.
The resulting inhomogeneous broadening of the CESR line gives
$1/\gamma \Delta H_{\text{inhom}}=T_{2}^{\ast }\ll T_{1,2}$,
and $T_{1}$ cannot be measured from the line-width. In Fig. \ref%
{ESR_linewidths} we show this effect: above $T_{c}$ MgB$_{2}$ has a
relaxationally broadened line-width of $\Delta H=0.9$ mT. Between
$T_{c}$ and the irreversibility temperature at the given field,
$T_{\text{irr}}$, the CESR remains homogeneous and narrows with the
lengthening of $T_{1}$. However, below $T_{\text{irr}}$ it broadens
abruptly and the line-width depends on the direction of the magnetic
field sweep: for up sweep it is broader than for down sweeps due to
the irreversibility of vortex insertion and removal.

To enable a direct measurement of the $T_1$ spin lattice relaxation time,
one has to resort to time resolved experiments. Conventional spin-echo ESR
methods are limited to $T_1$'s larger than a few 100 ns. To measure $T_1$'s
of a few nanoseconds, the so-called longitudinally detected ESR was invented
in the 1960's by Herv\'e and Pescia \cite{hervepescia} and improved by
several groups \cite{Atsarkin,Schweiger}. The method is based on the deep
amplitude modulation of the microwave excitation with an angular frequency, $%
\Omega \sim 1/T_1$. When the sample is irradiated with the amplitude
modulated microwaves at ESR resonance, the component of the
magnetization along the static magnetic field, $M_z$, decays from
the equilibrium value, $M_0$, with a time constant $T_1$. $M_z$
relaxes back to $M_0$ with a $T_1$ relaxation time when the
microwaves are turned off. The oscillating $M_z$ is detected using a
coil which is part of a resonant \textit{rf} circuit. The phase
sensitive detection of the oscillating voltage using lock-in
detection allows the measurement of $T_1$ using $\Omega T_1=v/u$
\cite{hervepescia,MuranyiJMR}, where $u$ and $v$ are the amplitudes
of the in- and out-of-phase components of the oscillating
magnetization after corrections for instrument related phase shifts.
The principal limitation of the LOD-ESR technique is its 3-4 orders
of magnitude lower sensitivity compared to conventional cw-ESR. The
LOD-ESR
method and the experimental apparatus are detailed in Refs. \cite%
{MuranyiJMR,SimonJMR}.

\begin{figure}[tbp]
\includegraphics[width=0.8\hsize]{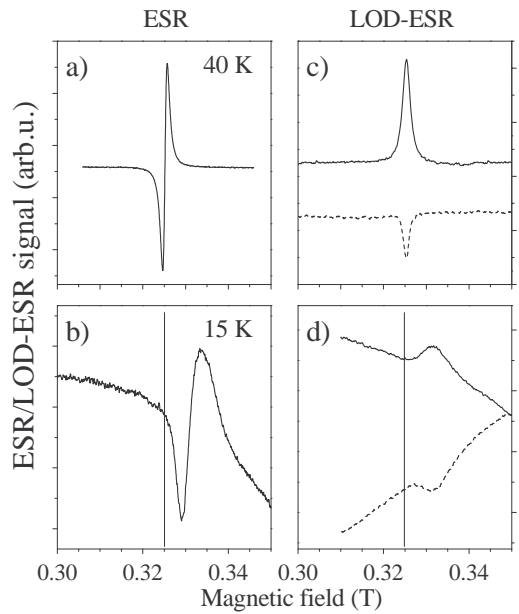}
\caption{ESR (a-b) and LOD-ESR (c-d) spectra of MgB$_2$ at 9.1 GHz
(0.32 T). a) and c) at 40 K in the normal state, and b) and d) in
the superconducting state at 15 K. Solid and dashed curves are the
in- and out-of-phase LOD signals, respectively and are offset for
clarity. Vertical solid lines indicate the resonance field above
$T_c$. Note the diamagnetic shift and broadening for for both kinds
of spectra below $T_c$. Also note the rotated phase of the in-phase
and out-of phase channels upon cooling. } \label{ESR_spectra}
\end{figure}

To prove that the LOD-ESR signal of the itinerant electrons is
detected in the superconducting phase, we compare in Fig.
\ref{ESR_spectra} the LOD-ESR signal with that measured
with conventional continuous-wave CESR (referred to as CESR in the following) of MgB$%
_2$ in the normal and superconducting states. The CESR signal is the
derivative of the absorption due to magnetic field modulation used
for lock-in detection. This signal was previously identified as the
ESR of conduction electrons in MgB$_2$ in the superconducting and
normal states \cite{SimonPRL2001,RettoriPRL2002,SimonMgB2PRB2005}
and its characteristics
have been discussed in detail \cite{SimonPRL2001,SimonMgB2PRB2005}. Above $%
T_c$ at 40 K, the CESR line is relaxationally broadened. Below
$T_c$, it is inhomogeneously broadened and is diamagnetically
shifted, i.e. to higher resonance fields. The irreversible effects
also contribute to a non-linear
baseline known as the non-resonant microwave absorption \cite{MullerPRB1987}%
. The intensity of the CESR signal decreases below $T_c$ as we
discussed previously \cite{SimonMgB2PRB2005}, due to the vanishing
of normal state electrons.

The LOD-ESR signal shows the same characteristics as the CESR below
$T_c$: it is broadened, shifted to higher fields and its intensity
decreases. The values for the temperature dependent diamagnetic
shifts and broadening and the relative intensity change agree for
the two kinds of measurements within experimental precision (not
shown). This unambiguously proves that the LOD-ESR signal originates
from the conduction electrons.

\begin{figure}[tbp]
\includegraphics[width=0.8\hsize]{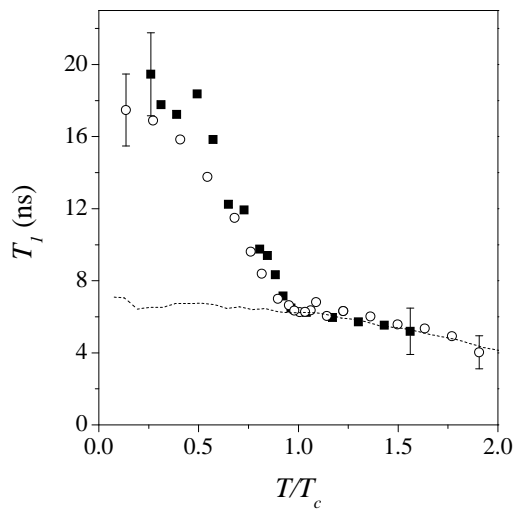}
\caption{Spin-lattice relaxation time as a function of the reduced
temperature in MgB$_2$ at 0.32 ($\blacksquare$) and 1.27 T
($\bigcirc$) magnetic fields. Representative error bars are shown
for some of the data. Dashed curve shows $T_1$ corresponding to
$\Delta H_{\text{Hom}}$ in the 15 T measurement such as in Fig.
\protect\ref{15T_Tdep} with the reduced temperature normalized to 39
K.} \label{T1_data}
\end{figure}

The change of the relaxation time $T_1$ is visible in the LOD-ESR
spectra in Fig. \ref{ESR_spectra} as a change in the relative
intensities of the in- and out-of-phase signals. At 40 K $v/u=0.47$
and at 15 K $v/u=0.95$, which together with $\Omega/2 \pi= 11.4
\text{ MHz}$ gives 6.3 and 13.3 ns relaxation times, respectively.
In Fig. \ref{T1_data}, we show the $T_1$ data inferred from the
LOD-ESR spectra at 0.32 and 1.27 T as a function of the reduced
temperature $T/T_c$.

\section{Discussion}

The observed lengthening of $T_{1}$ below $T_{c}$ ( Fig.
\ref{T1_data}) is expected from theory for non-magnetic scattering
centers and low magnetic fields where the susceptibility is
dominated by excitations over the superconducting gap. On the other
hand, the field independence between 0.32 and 1.27 T of $T_{1}$
below $T_{c}$ is surprising. The lengthening of $T_{1}$ below
$T_{c}$ in zero magnetic field for an isotropic, type I
superconductor
was calculated in the framework of weak-coupled BCS theory by Yafet \cite%
{Yafet}. He concluded that $T_{1}$ lengthens as a result of the
freezing of normal state excitations. However, no theory exists for
a type II superconductor in finite fields with $H_{c2}$ anistotropy
such as MgB$_{2}$, thus here the $T_{1}$ data are analyzed
phenomenologically in the framework of the two-band/gap model of
MgB$_{2}$.

In the following, we deduce the residual (low temperature), impurity
related spin scattering contributions of the $\sigma $ and $\pi $\
Fermi surface sheets. The DOS is distributed almost equally on the
FS sheets of MgB$_{2}$: $N_{\pi }/(N_{\pi }+N_{\sigma })=0.56$
\cite{LouieNAT2002}, where $N_{\pi }$ and $N_{\sigma }$ are the DOS
of the two types of FS sheets. A magnetic field of $\sim 0.3-0.4$ T
closes the gap on the $\pi $ FS sheets but leaves the gap on the
$\sigma $ sheet almost intact.
\cite{BouquetPRL2002,SimonMgB2PRB2005}. This suggests that well
below $T_{c}$, our experiment at 0.32 T measures exclusively the
relaxation of electrons on the fully closed $\pi $ FS sheets. Since
$T_{1}$ at 0.32 T increases slowly with temperature between 10 and
20 K, we extrapolate $T_{1\pi }\approx T_{1}(\text{10 K, 0.32
T})=20(2)$ ns for the $\pi $ FS.

In order to separate the contribution of the $\sigma $ FS to the
relaxation rate in the normal state, $1/T_{1n}$,  we assume that
inter-band relaxation is negligible and $1/T_{1n}$ is equal to the
average of the spin-lattice relaxation rates on the two FS's
weighted by the corresponding DOS:

\begin{equation}
\frac{1}{T_{1n}}=\frac{N_{\pi }/T_{1\pi}+N_{\sigma }/T_{1\sigma
}}{N_{\pi }+N_{\sigma }}  \label{ComposedRelaxation}
\end{equation}

\noindent Here $T_{1\sigma }$ is the spin-lattice relaxation time on
the $\sigma$ FS. The 15 T measurement shows that $1/T_{1n}$ changes
little with temperature between 3 K and 40 K. Thus we find
$T_{1\sigma }=3.4(5)$ ns for the contribution of the $\sigma $ FS
sheets using $T_{1n}=T_{1}(T_{c})=6.3$ ns, $T_{1\pi }=20(2)$ ns and
Eq. \ref{ComposedRelaxation}.

For normal metals with a simple Fermi surface, the so-called Elliott
relation \cite{Elliott,Yafet,YafetSSP1963,BeuneuMonodPRB1977} holds, which
states that for a given type of disorder (e.g. phonons or dislocations) $%
T_{1}$ is proportional to the momentum relaxation time, $\tau $. The
proportionality constant depends on the spin orbit splitting of the
conduction electron bands and has been estimated in a number of
metals from the shift of the CESR from the free electron value.
Metals with complicated Fermi surfaces i.e. with great variations of
the electron-phonon coupling on the different FS parts are known to
deviate from the Elliott relation \cite{MonodBeuneuPRB1979} and
calculation of $T_{1}$ requires to take into account the details of
the band-structure
\cite{SilsbeeBeneuPRB1983,FabianPRL1998,FabianPRL1999}. Examples
include polyvalent elemental metals such as Mg or Al. Clearly, a
calculation of $T_{1}$ is required for MgB$_{2}$, which takes into
account its band structure peculiarities. Comparison of spin
scattering and momentum scattering times of the two types of Fermi
surfaces is instructive. The relative values of $\tau $ for the two
FS parts, $\tau _{\pi }$ and $\tau _{\sigma }$, and for
interband scattering were estimated by Mazin \textit{et al.} \cite%
{MazinPRL2002}. A very small impurity interband scattering and
$\tau_{\pi}<\tau_{\sigma}$, i.e. a larger $\pi $ intraband
scattering relative to $\sigma $ intraband scattering was required
to explain the rather small depression of T$_{c}$ in materials with
widely different conductivities.  De Haas-van Alphen
\cite{CooperPRL2002} and magnetoresistance \cite{PallecchiPRB2005}
measurements of high purity samples yield $\tau _{\pi }<\tau
_{\sigma }$ also. Such a behaviour could rise from Mg vacancies,
which perturb more electrons of the $\pi $ band relative to the
$\sigma $ band. However, our spin scattering data do not support
this. In contrast to momentum scattering, spin scattering is
stronger on the $\sigma$ FS: $T_{1\pi }:T_{1\sigma }=6:1$  in high
purity samples and low temperatures. The relative values of $T_{1}$
and $\tau $ for the two FS do not necessarily follow the same trend,
spin relaxation times at low temperatures depend on spin orbit
relaxation on impurities while momentum relaxation is due to
potential scattering.  However, a defect center such as a Mg vacancy
with a strong modification of the electron-phonon coupling and an
atomic number strongly differing from that of the regular atoms
constituting the compound would greatly affect $T_{1}$ compared to
$\tau $. In the two gap model Mg defects are expected to shorten
$T_{1\pi }$ more than $T_{1\sigma }$ and thus are unlikely to be the
dominant scatterers.

A final note concerns the validity of the above analysis of
$T_{1}$'s in the framework of the two-band/gap model. The field
independence of the lowest temperature $T_{1}$ for 0.32 and 1.27 T
is unexpected within this model. The spin susceptibility increases
strongly between these fields and more normal states are restored at
1.27 T than expected from the closing of the gap on the $\pi $ FS
sheets alone \cite{SimonMgB2PRB2005}. Based on this, one would
expect to observe additional spin scattering from the restored
$\sigma $ FS parts, which is clearly not the case. This also
indicates that a theoretical study, which takes into account the
peculiarities of MgB$_{2}$ is required to explain the anomalous
spin-lattice relaxation times.

In conclusion, we presented the measurement of the spin-lattice
relaxation time, $T_{1}$, of conduction electrons as a function of
temperature and magnetic field in the MgB$_{2}$ superconductor. We
use a novel method based on the detection of the $z$ component of
the conduction electron magnetization during electron spin resonance
conditions with amplitude modulated microwave excitation.
Lengthening of $T_{1}$ below $T_{c}$ is observed irrespective of the
significant CESR line broadening due to irreversible diamagnetism in
the polycrystalline sample. The field independence of $T_{1}$ for
0.32 T and 1.27 T allows to measure the separate contributions to
$T_{1}$ from the two distinct types of the Fermi surface.


\section{Acknowledgements}
The authors are grateful to Rich\'ard Ga\'al for the development of
the ESR instrument at the EPFL. F.S. and F.M. acknowledge the
Zolt\'{a}n Magyary postdoctoral programme, the Bolyai fellowship of
the Hungarian Academy of Sciences and the Alexander von Humboldt
Foundation for support, respectively. Work supported by the
Hungarian State Grants (OTKA) No. TS049881, F61733, PF63954 and
NK60984 and by the Swiss NSF and its NCCR "MaNEP". Ames Laboratory
is operated for the U.S. Department of Energy by Iowa State
University under Contract No. W-7405-Eng-82.

$^{\ast }$ Corresponding author: simon@esr.phy.bme.hu

$^{\dag }$ Present address: Leibniz Institute for Solid State and Materials
Research Dresden, PF 270116 D-01171 Dresden, Germany

$^{\ddag}$ Present address: Condensed Matter Physics and Materials
Science Department, Brookhaven National Laboratory, Upton, New York
11973-5000, USA


\end{document}